\documentclass[pdflatex,sn-aps]{sn-jnl}

\usepackage{graphicx}%
\usepackage{multirow}%
\usepackage{amsmath,amssymb,amsfonts}%
\usepackage{amsthm}%
\usepackage{mathrsfs}%
\usepackage[title]{appendix}%
\usepackage{xcolor}%
\usepackage{textcomp}%
\usepackage{manyfoot}%
\usepackage{booktabs}%
\usepackage{algorithm}%
\usepackage{algorithmicx}%
\usepackage{algpseudocode}%
\usepackage{listings}%
\usepackage{hyperref}
\usepackage{url}

\theoremstyle{thmstyleone}%
\theoremstyle{thmstyletwo}%

\theoremstyle{thmstylethree}%

\begin{document}

\title[Article Title]{An investigation of phrase break prediction in an End-to-End TTS system}


\author*{\fnm{Anandaswarup} \sur{Vadapalli}}\email{anandaswarup.vadapalli@research.iiit.ac.in}



\affil[]{\orgdiv{Speech Processing Lab, Language Technologies Research Center},
    \orgname{International Institute of Information Technology, Hyderabad},
    \orgaddress{\city{Hyderabad}, \postcode{500032}, \state{Telangana}, \country{India}}}

\abstract{\textbf{Purpose:} This work explores the use of external phrase break prediction models to enhance listener comprehension in End-to-End Text-to-Speech (TTS) systems.
    
    \textbf{Methods:} The effectiveness of these models is evaluated based on listener preferences in subjective tests. Two approaches are explored: (1) a bidirectional LSTM model with task-specific embeddings trained from scratch, and (2) a pre-trained BERT model fine-tuned on phrase break prediction. Both models are trained on a multi-speaker English corpus to predict phrase break locations in text. The End-to-End TTS system used comprises a Tacotron2 model with Dynamic Convolutional Attention for mel spectrogram prediction and a WaveRNN vocoder for waveform generation
    
    \textbf{Results:} The listening tests show a clear preference for text synthesized with predicted phrase breaks over text synthesized without them.
    
    \textbf{Conclusion:} These results confirm the value of incorporating external phrasing models within End-to-End TTS to enhance listener comprehension.}


\maketitle

\titlepage

\section{Introduction}
\label{sec:introduction}

Text-to-Speech (TTS) systems face the challenge of reconstructing audio from compressed textual input. Unlike machine translation, which involves sequences of similar lengths, TTS generates much longer audio sequences from short text prompts. The complexity is further increased by the numerous ways a single sentence can be spoken.

In recent years, deep learning architectures based on sequence-to-sequence (Seq2Seq) models have gained traction in TTS due to their ability to learn from text-audio data pairs without the need for intricate sub-systems. However, these models struggle to explicitly control stylistic aspects like prosody and speaking style, as they inherently integrate out such unlabeled attributes during training. To address this limitation, researchers have developed methods to control these latent attributes within End-to-End TTS frameworks~\citep{Skerry-Ryan2018, Wang2018, Zhang2019, Lee2019, Hsu2019, Battenberg2019, Aggarwal2020, Valle2020, Hu2020, DBLP:journals/spl/LiuSBGL20, DBLP:conf/icassp/XiaoHMS20}.

This paper investigates the role of phrase break prediction in End-to-End TTS systems. We explore the following key questions:

\begin{enumerate}
    \item \textbf{Does including an external, explicit phrasing model benefit End-to-End TTS?} We hypothesize that incorporating a dedicated model for predicting phrase breaks can improve the system's capability to generate natural-sounding speech.
    \item \textbf{How can we assess the effectiveness of an external phrasing model?} We propose using listener comprehension to evaluate the model's impact on the clarity and understandability of synthesized speech. Our core hypothesis is that employing an external phrasing model within an End-to-End TTS system will lead to enhanced listener comprehension of the synthesized speech.
\end{enumerate}

To address these questions, we design a framework that integrates and evaluates external phrase break prediction models in End-to-End TTS systems. Specifically, we introduce two external phrase break prediction models, (i) a BLSTM model with task-specific embeddings, and (ii) a fine-tuned BERT model, both aimed at enhancing the naturalness and listener comprehension of synthesized speech. Using a robust evaluation framework, we demonstrate that both models significantly improve listener comprehension, with the BERT model outperforming the BLSTM model in both objective metrics and subjective listening tests. Furthermore, the proposed approach highlights the importance of integrating explicit phrasing mechanisms into TTS systems for tasks like storytelling, where prosodic clarity plays a critical role. These results validate our hypothesis regarding the utility of external phrasing models in End-to-End TTS systems.

This paper is structured as follows. Section~\ref{sec:phrase_break} provides a background on phrase break prediction for TTS. Section~\ref{sec:phrasing_models} details the phrasing models employed in this work. Section~\ref{sec:end_to_end_tts} describes the End-to-End TTS system we utilize. Section~\ref{sec:end_to_end_tts_phrasing} addresses the integration and evaluation of our phrasing models within the TTS system. Finally, Section~\ref{sec:summary_conclusions} summarizes the paper and presents our conclusions.

\section{Phrase break prediction}
\label{sec:phrase_break}

Spoken language has a natural structure where words group together, separated by brief pauses or disjunctures. This organization can be described using the concept of prosodic phrasing, analogous to the syntactic structure of written sentences.

Phrase breaks serve a critical purpose in natural speech. Physiologically, they provide brief pauses during speech. More importantly, they emphasize content and enhance speech intelligibility. These breaks can be categorized into various levels~\cite{Silverman1992}. Notably, they exhibit a non-linear relationship with syntactic breaks~\cite{taylor1998} and vary depending on the speaker~\cite{kishoreSSW7, kishore2010}. Additionally, the phrasing style is influenced by the text type itself. For instance, the phrasing pattern used for reading news may differ from that employed for storytelling.

The process of introducing phrase breaks within an utterance is termed phrasing. In the context of speech synthesis, phrasing plays a crucial role. By segmenting lengthy utterances into meaningful information units, phrasing significantly improves the synthesized speech's intelligibility.  Furthermore, it often serves as the foundation for other prosody models, including those for accent prediction, intonation prediction~\cite{Hirschberg1993, Ross1996}, and duration modeling~\cite{Santen1994}. Errors introduced during the initial phrasing stage can propagate to subsequent prosody models, ultimately leading to unnatural and difficult-to-understand synthesized speech.

\subsection{Phrase break prediction in TTS systems}
\label{subsec:phrase_break_tts}

Several acoustic cues in the speech signal indicate phrase breaks. These include relative changes in intonation and syllable duration, as well as pauses. Additional cues, such as pre-pausal lengthening of rhymes, speaking rate, breaths, boundary tones, and glottalization, also contribute to indicating phrase breaks~\cite{wightman1992, redi2001, kim2006}. However, representing these non-pausal cues through features remains a challenge due to the complexities involved~\cite{kishoreSSW7}. Consequently, most prior works, including this one, focus on predicting pause locations during speech synthesis~\cite{parlikar2013, wattsBlizzard2013, anandaswarup2013, watts2011, anandaswarupIS14, wattsICASSP2014, anandaswarupIS16}.

Given this context, the phrase break prediction task within TTS can be formally defined as follows: ``\textit{Given an utterance represented as a sequence of words intended for TTS synthesis, predict at each word boundary whether a pause should be inserted after that word}."

Traditionally, machine learning models like regression trees or Hidden Markov Models (HMMs) were employed for phrase break prediction. These models relied on data labeled with linguistic features such as part-of-speech (POS) tags and phrase structure information~\cite{parlikar2011, wang1992, navas2008, taylor1998, schmid2004, bonafonte2004, busser2001}. Additionally, significant research focused on unsupervised methods for inducing word representations that could serve as substitutes for POS tags or other linguistic classes within the phrase break prediction task~\cite{parlikar2012, hema2004, anandaswarup2013}.

The emergence of deep learning based techniques for generating continuous word representations, known as word embeddings, led to their application in phrase break prediction~\cite{wattsBlizzard2013, watts2011, wattsThesis, anandaswarupIS14, wattsICASSP2014}. In~\cite{anandaswarupIS16}, the authors modeled phrase break prediction as a sequence modeling task and demonstrated that recurrent neural network (RNN) models outperform feedforward deep neural network (DNN) models for this purpose. In~\cite{DBLP:conf/interspeech/KlimkovNMPBMD17} the authors experiment with different textual features using both classification and regression trees (CART) and bidirectional long short term memory (BLSTM) models, and conclude that using a BLSTM model with word embeddings proves to be beneficial for phrase break prediction. In~\cite{DBLP:conf/icassp/XiaoHMS20}, the authors use linguistic information as well as a bidirectional encoder representations from transformers (BERT) language model~\cite{DBLP:conf/naacl/DevlinCLT19} to improve the prosody of their Mandarin TTS system. In~\cite{DBLP:journals/spl/LiuSBGL20}, the authors implemented multi-task learning to learn a word-level prosody embedding as a secondary task, which was subsequently used to model prosodic phrasing within a Tacotron-based TTS system. In~\cite{DBLP:conf/interspeech/FutamataPYT21}, the authors explored combining representations from a BLSTM language model with a fine-tuned BERT model for phrase break prediction in Japanese TTS synthesis.

This work distinguishes itself from prior research by integrating phrase break prediction models into End-to-End TTS systems, emphasizing their role in enhancing listener comprehension. While earlier studies, such as~\cite{anandaswarupIS16, DBLP:conf/icassp/XiaoHMS20}, have utilized BLSTM and BERT models independently for phrase break prediction, our approach goes further by directly comparing the effectiveness of a task-specific BLSTM model and a fine-tuned BERT model within a complete synthesis pipeline. Moreover, unlike~\cite{DBLP:conf/interspeech/FutamataPYT21}, which combined representations for Japanese TTS, our study employs a robust subjective evaluation methodology with multi-speaker English datasets and real-world use cases like children's storytelling, demonstrating the broader applicability of these models. These results validate our hypothesis regarding the utility of external phrasing models in End-to-End TTS systems.


\section{Phrasing models}
\label{sec:phrasing_models}

\subsection{Data used}
\label{subsec:phrasing_models_data}

Our experiments use the LibriTTS dataset~\cite{DBLP:conf/interspeech/ZenDCZWJCW19}, a multi-speaker English corpus of approximately 585 hours of read English speech designed for TTS research. We trained our phrasing models using the publicly available LibriTTS alignments, created using the Montreal-Forced-Aligner~\cite{DBLP:conf/interspeech/McAuliffeSM0S17} with the pre-trained English model. These alignments provide precise pause locations, serving as ground truth for training the phrasing models. For training, we employed the `train-clean-360' split, while `dev-clean' and `test-clean' splits were reserved for validation and testing, respectively.

\subsection{Systems built}
\label{subsec:phrasing_models_systems}

This section details the two phrase break prediction models we train on the dataset described in Section~\ref{subsec:phrasing_models_data}.

\begin{enumerate}
    \item \textbf{Model 1: BLSTM with task-specific embeddings}: The first model is a BLSTM network, that uses task-specific static word embeddings trained from scratch. This model serves as our baseline and shares similarities with the model presented in~\cite{anandaswarupIS16}. This model is henceforth referred to as the BLSTM Model.
    \item \textbf{Model 2: BERT model fine-tuned on phrase break prediction}: The second model consists of a pre-trained BERT model fine-tuned for phrase break prediction. This model is henceforth referred to as the BERT Model.
\end{enumerate}

We describe both models in detail below. 

\subsubsection{Model 1: BLSTM with task-specific embeddings (BLSTM Model)}
\label{subsubsec:phrasing_models_systems_blstm}

\begin{figure*}[!htb]
    \centering
    \includegraphics[scale = 0.4]{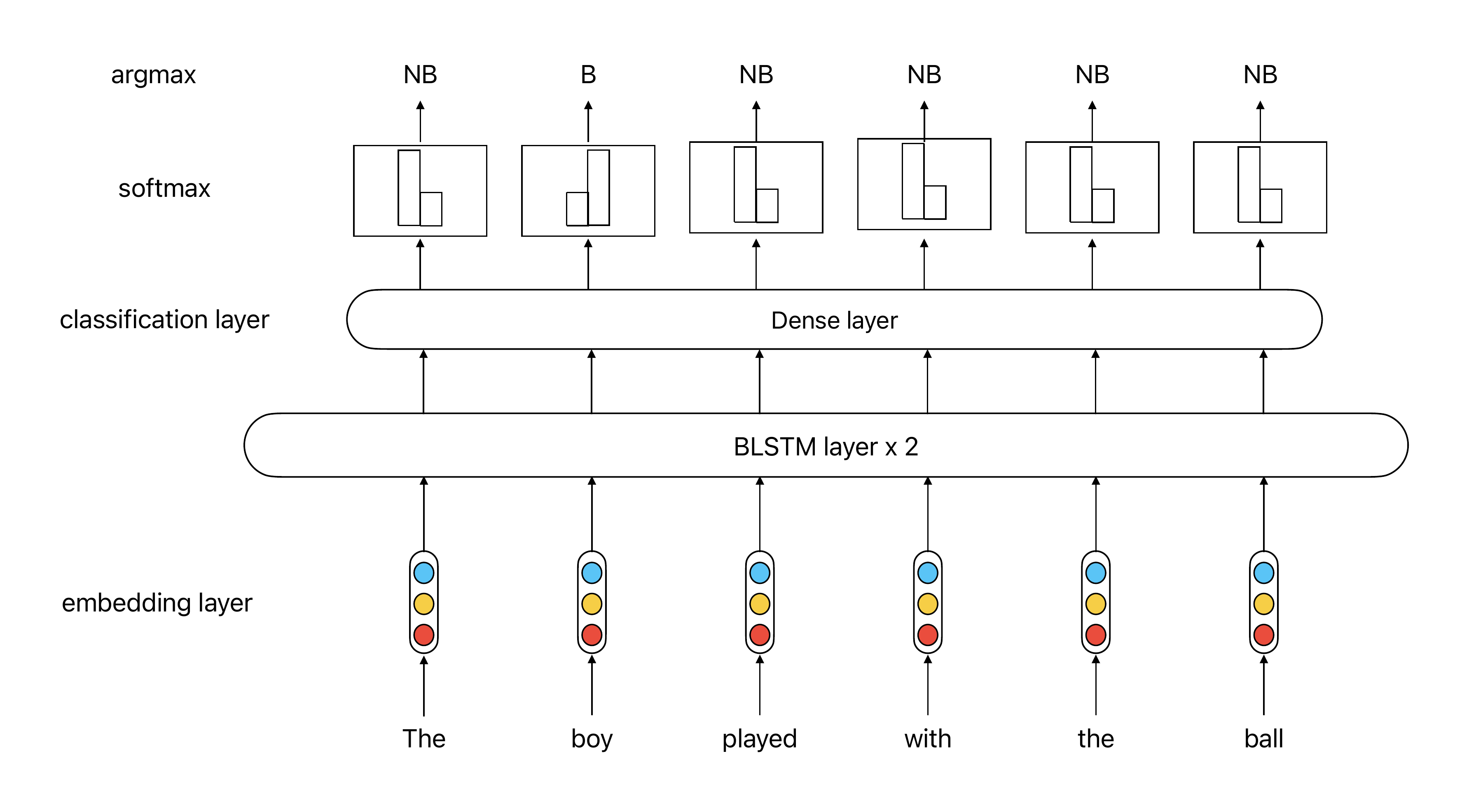}
    \caption{BLSTM token classification model using task-specific static embeddings trained from scratch. Inputs to the model are word embeddings which are randomly initialized and jointly trained along with the model on the task at hand, outputs of the model are probabilities from a softmax layer over the set of possible tags (B or NB).}
    \label{fig:blstm_token_classification_model}
\end{figure*}

This section details our first phrasing model, a BLSTM network for token classification. Given an input sequence of words, the model predicts for each word boundary whether a pause (break) should be inserted afterward. The model architecture consists of three main components, as shown in Figure~\ref{fig:blstm_token_classification_model}:

\begin{enumerate}
    \item \textbf{Word Embeddings}: Each word in the input sequence is mapped to a corresponding word embedding. These embeddings are randomly initialized and jointly trained along with the model on the task at hand i.\thinspace e. phrase break prediction.
    \item \textbf{BLSTM Layers}: The sequence of word embeddings is then fed into a stack of two BLSTM layers. These layers capture contextual information within the sequence.
    \item \textbf{Binary Classifier}: The output from the stacked BLSTM layers, representing each input token, is fed into a simple binary classifier. This classifier consists of a dense feed-forward layer followed by a softmax activation function. The softmax layer outputs probabilities for each word boundary belonging to the set of possible break labels (Break ``B" or No Break ``NB").
\end{enumerate}

The model was randomly initialized and trained to convergence using the Adam optimizer~\cite{DBLP:journals/corr/KingmaB14} and a cross-entropy loss. Table~\ref{tab:blstm_token_classification_model_parameters} shows the model parameters as well as training hyperparameter values.

\begin{table}[!htb]
    \caption{Model parameters and training hyperparameter values for BLSTM Model}
    \centering
    \begin{tabular}{|c|c|} \hline
        word embedding dimension  & 300                                      \\ \hline
        number of BLSTM layers    & 2                                        \\ \hline
        BLSTM hidden layer size   & 512                                      \\ \hline
        training batch size       & 64                                       \\ \hline
        training optimizer        & Adam~\cite{DBLP:journals/corr/KingmaB14} \\ \hline
        learning rate             & 0.001                                    \\ \hline
        number of training epochs & 10                                       \\ \hline
    \end{tabular}
    \label{tab:blstm_token_classification_model_parameters}
\end{table}

During inference, to generate a sequence of break indices for a given text input, a greedy approach is followed. We first run the forward pass of the model on the input sequence, and select the most likely break index (B or NB) for each input token by applying an argmax over the probabilities generated by the softmax layer corresponding to that particular token.

\subsubsection{Model 2: BERT model fine-tuned on phrase break prediction (BERT Model)}
\label{subsubsec:phrasing_models_systems_blstm_bert}

\begin{figure*}[!htb]
    \centering
    \includegraphics[scale = 0.4]{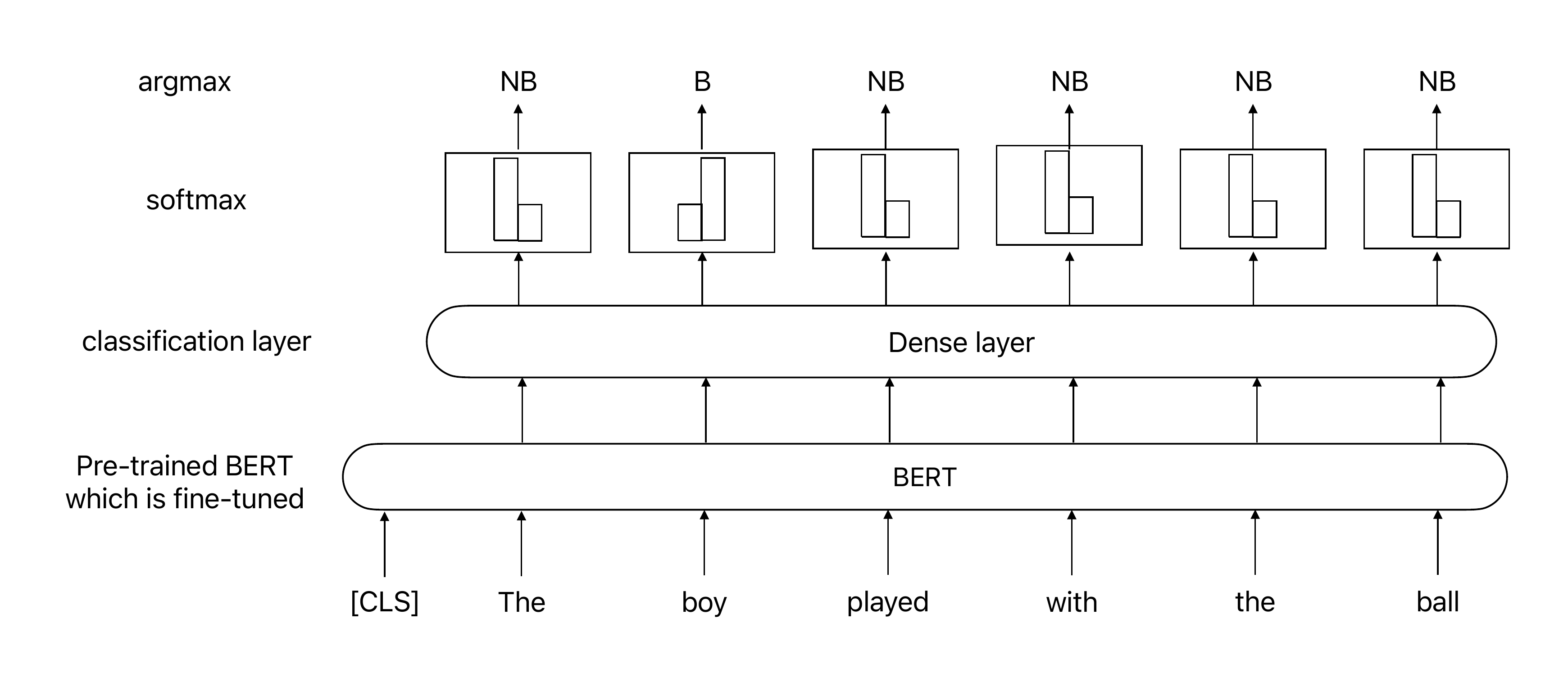}
    \caption{BERT model with a token classification head. The BERT model was pre-trained on uncased English text and was later fine-tuned on phrase break prediction. The classification layer was randomly intialized and it's parameters were learnt from scratch.}
    \label{fig:bert_token_classification_model}
\end{figure*}

This model consists of a pre-trained BERT model, with an additional token classification layer to perform phrase break prediction. As we had neither the data nor the computational resources to train a BERT model from scratch, we made use of the Transformers\footnote{\url{https://huggingface.co/docs/transformers/index}} library which provides APIs and tools to easily download and fine-tune state-of-the-art pre-trained models. We used the `bert-base-uncased' model from the Transformers library, which is a pre-trained $\text{BERT}_{BASE}$ model trained on uncased English text. Figure~\ref{fig:bert_token_classification_model} shows the architecture of this model.

The classification layer was randomly initialized and the entire model was fine-tuned on phrase break prediction using the Adam optimizer~\cite{DBLP:journals/corr/KingmaB14} and a cross-entropy loss. During training, the parameters of the classification layer were learnt from scratch while the parameters of the pre-trained BERT model were fine-tuned on phrase break prediction. Table~\ref{tab:bert_finetuning_parameters} shows the fine-tuning parameter values used for this model.

Similar to the BLSTM model, the BERT model follows a greedy approach during interence to generate a sequence of break indices for a given text input. We first run the forward pass of the model on the input sequence, and select the most likely break index (B or NB) for each input token by applying an argmax over the probabilities generated by the softmax layer corresponding to that particular token.


\begin{table}[!htb]
    \caption{Fine-tuning hyperparameter values for BERT Model}
    \centering
    \begin{tabular}{|c|c|} \hline
        bert model name        & `bert-base-uncased' \\ \hline
        batch size             & 64                  \\ \hline
        learning rate          & 0.00001             \\ \hline
        gradient clipping norm & 10                  \\ \hline
        number of epochs       & 10                  \\ \hline
    \end{tabular}
    \label{tab:bert_finetuning_parameters}
\end{table}

\subsection{Objective evaluation of phrasing models}
\label{subsec:phrasing_models_objective_results}

Table~\ref{tab:bidirectional_encoder_speaker_independent_results} shows the performance of both models described in Section~\ref{subsec:phrasing_models_systems} on the phrase break prediction task. We report our results in terms of the F1 Score~\cite{FMeasure} which is defined as the harmonic mean of precision and recall.

\begin{table}[!htb]
    \caption{Performance (in terms of the F1 Score) of the BLSTM Model and the BERT model on phrase break prediction}
    \centering
    \begin{tabular}{|c|c|c|} \hline
        BLSTM Model & 88.91          \\ \hline
        BERT Model  & \textbf{92.10} \\ \hline
    \end{tabular}
    \label{tab:bidirectional_encoder_speaker_independent_results}
\end{table}

An examination of the results shows that the BERT model outperforms the BLSTM model on phrase break prediction.

\section{End-to-End TTS system}
\label{sec:end_to_end_tts}

The End-to-End speech synthesis system consists of:

\begin{enumerate}
    \item A Tacotron2 model with Dynamic Convolutional Attention, which modifies the hybrid location sensitive attention mechanism to improve generalization on long utterances. This model takes text as input and predicts a sequence of mel spectrogram frames as output.
    \item A WaveRNN based vocoder, which generates a waveform from the mel spectrogram predicted by the Tacotron2 model.
\end{enumerate}

\subsection{Architecture Details}
\label{subsec:end_to_end_architecture}

We describe the architecture details of both the Tacotron2 and WaveRNN models below. The mel spectrograms used for training both models are computed from 22,050 Hz audio using a 50 ms frame size, a 12.5 ms frame shift, an FFT size of 2048 and a Hann window. The FFT energies are then transformed to the mel scale using an 80 channel mel filterbank followed by log dynamic range compression.

\subsubsection{Tacotron2 model}
\label{subsubsec:end_to_end_tacotron2}

\begin{figure*}[!htb]
    \centering
    \includegraphics[scale=0.4]{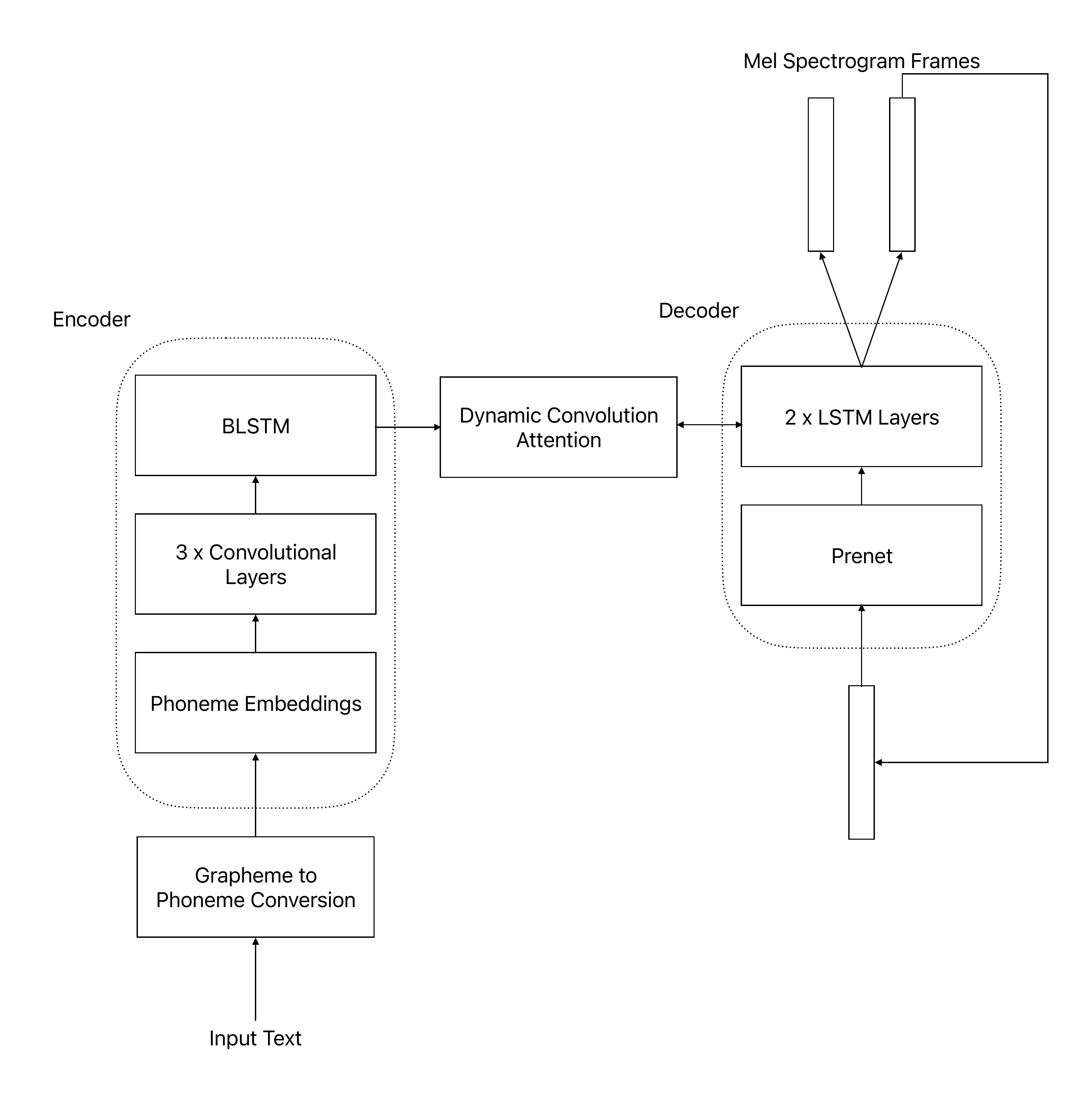}
    \caption{Architecture of the Tacotron2 model. The Tacotron2 model takes text as input and predicts a sequence of mel spectrogram frames as output.}
    \label{fig:tacotron2}
\end{figure*}

The Tacotron2 model used in this work is based on the system described in~\cite{DBLP:conf/icassp/BattenbergSMSKS20}, which is composed of an encoder and decoder with attention. Figure~\ref{fig:tacotron2} shows the block diagram of this model. We use phoneme sequence in combination with punctuation and word boundaries as input to the Tacotron2 model. This way mispronunciations are reduced and the network learns appropriate pausing through the punctuation. More specifically, the Tacotron2 model learns to insert a pause whenever it encounters a comma in the text.

The encoder takes an input text sequence and maps this into a sequence of hidden states. The input phonemes are represented using a learned $512$ dimensional embedding., which is then passed through a stack of 3 convolutional layers, each containing $512$ filters of shape $5 \times 1$ followed by batch normalization and ReLU activation. These convolutional layers model longer-term context in the input text sequence. The output of the final convolutional layer is passed to a single bidirectional LSTM layer of $512$ units ($256$ forward units + $256$ backward units) to generate the encoded hidden state sequence (also called the memory) which is the output of the encoder.

The output of the encoder is passed to the attention network which summarizes the full encoded sequence as a fixed-length context vector for each decoder output step. Our system uses dynamic convolution attention~\cite{DBLP:conf/icassp/BattenbergSMSKS20}, which uses 1-D convolutions consisting of 8 filters of length 21 for the static and dynamic filters respectively, and a 1-D convolution consisting of 1 filter of length 11 for the causal prior filter.

The decoder is an autoregressive recurrent neural network which predicts a sequence of mel spectrogram frames from the encoded input sequence, one frame at a time. The prediction from the previous timestep is passed to through a prenet containing 2 fully connected layers of 256 units with ReLU activations. This prenet acts as an information bottleneck and is essential for learning attention. The prenet output and the attention context vector are concatenated and passed through a stack of 2 LSTM layers of 1024 units each. The LSTM output is concatenated with the attention context vector and projected through a linear transform to predict the target mel spectrogram frame. We use a reduction factor of 2, i.\thinspace e. we predict two mel spectrogram frames for each decoder step.

All the convolutional layers in the network are regularized using dropout with probability $0.5$ while the LSTM layers are regularized using dropout with probability $0.1$. To ensure output variation at inference time, dropout with probability $0.5$ is applied to the layers of the decoder prenet.

\subsubsection{WaveRNN model}
\label{subsubsec:wavern_model}

\begin{figure*}[!htb]
    \centering
    \includegraphics[scale=0.4]{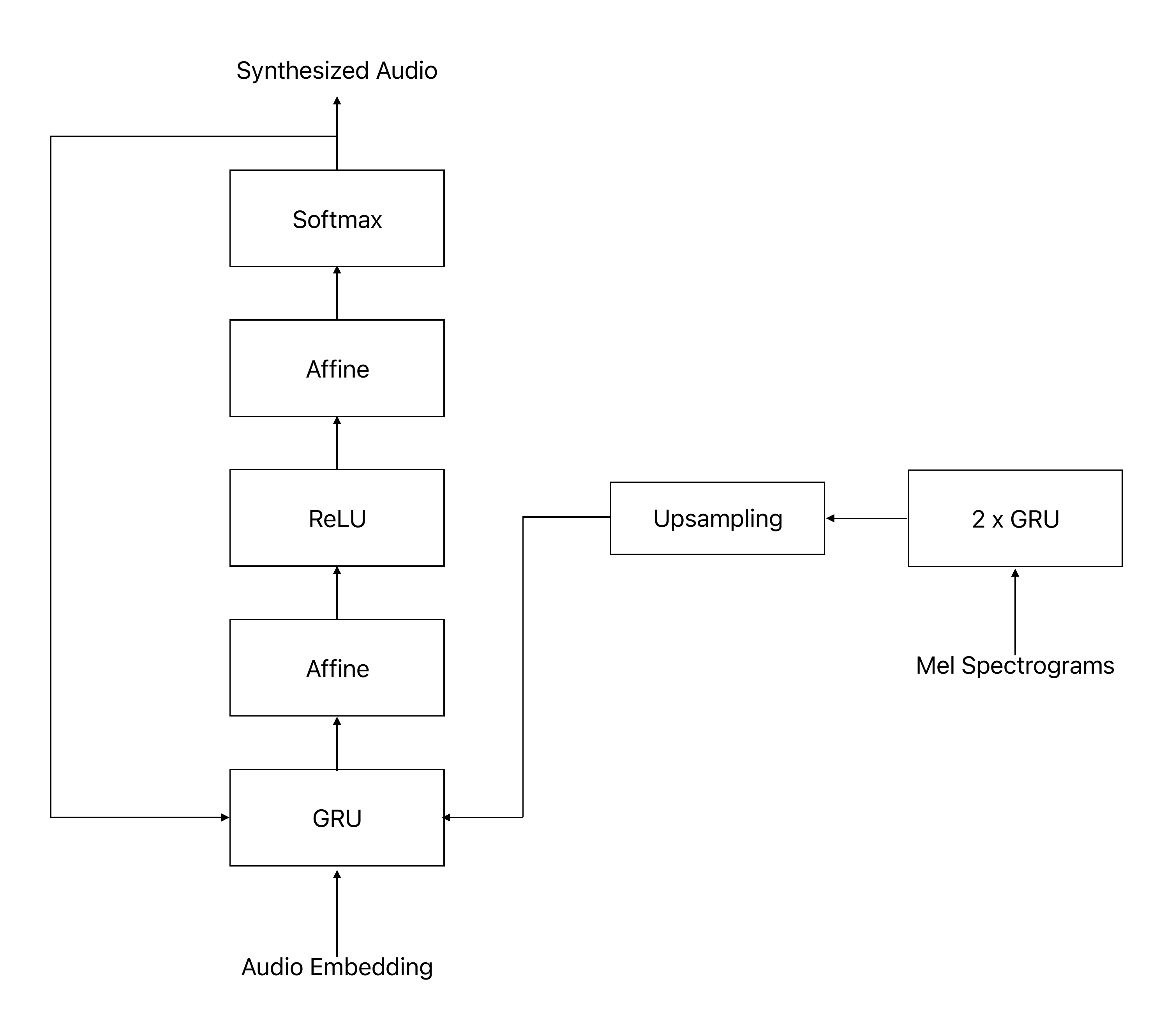}
    \caption{Architecture of the WaveRNN model. The WaveRNN model takes the mel spectrogram predicted by the Tacotron2 model as input and generates a waveform as output.}
    \label{fig:wavernn}
\end{figure*}

The WaveRNN model used in this work is based on the system described in~\cite{DBLP:conf/icml/KalchbrennerESN18, DBLP:conf/interspeech/Lorenzo-TruebaD19}, which consists of a conditioning network and an autoregressive network. The conditioning network consists of a pair of bidirectional GRU layers of $128$ units each. The autoregressive network is a single GRU layer of $896$ units followed by a pair of affine layers and finally a softmax layer of $1024$ units which predicts $10$ bit mu-law quantized audio samples. Figure~\ref{fig:wavernn} shows the block diagram of this model.

\subsection{Training details}
\label{subsec:training}

We train both the Tacotron2 and the WaveRNN models seperately on the LJSpeech dataset~\cite{ljspeech17}, a public dataset consisting of 13,100 short audio clips of a single speaker, reading passages from 7 non-fiction books. The total length of the dataset is $\sim$24 hrs. For both models, we train on a 12,838 utterance subset ($\sim$23 hrs), while our validation and held out test subsets have 131 utterances each.

During Tacotron2 training, we augment the original LJSpeech texts with commas where ever there is a pause in the corresponding audio clip. As described in Section~\ref{subsubsec:end_to_end_tacotron2} we then use the phoneme sequence in combination with the augmented commas and word boundaries as input to the Tacotron2 model. This is done so that the network learns to insert a pause in the synthesized speech whenever there is a comma in the text.

The Tacotron2 model is trained using teacher-forcing (the ground truth mel spectrogram is used as the input for each decoder step). We train using the Adam optimizer for $300$k steps with a batch size of $128$ on a single Nvidia GeForce RTX 2080 Ti GPU. We use an initial learning rate of $1 \times 10^{-3}$ which is then reduced to $5 \times 10^{-4}$, $2.5 \times 10^{-4}$, $1.25 \times 10^{-4}$, $6.25 \times 10^{-5}$ and $3.125 \times 10^{-5}$ at $20$k, $40$k, $100$k, $150$k and $200$k steps respectively.

We train the WaveRNN model using the Adam optimizer for $300$k steps with a batch size of $32$ on a single Nvidia GeForce RTX 2080 Ti GPU. We use an initial learning rate of $4 \times 10^{-4}$ which is halved every $25$k steps.

\section{Incorporation and evaluation of the phrasing models in the End-to-End TTS system}
\label{sec:end_to_end_tts_phrasing}

This section describes the evaluation of the two phrasing models (described in Section~\ref{subsec:phrasing_models_systems}), using listener comprehension as the primary metric. Our core hypothesis is that incorporating an external phrasing model within the End-to-End TTS system will lead to improved listener comprehension of the synthesized speech.

In written text, punctuation marks like commas typically indicate phrase breaks. Typically, TTS systems insert a pause in the synthesized speech whenever they encounter a comma in the text. As described in Section~\ref{subsec:training} the End-to-End TTS model used in this work is also trained to insert a pause whenever it encounters a comma in the text. When presented with unpunctuated text, both phrasing models described in Section~\ref{sec:phrasing_models} predict comma locations, essentially punctuating the text. The punctuated text is then passed to the TTS system for synthesis.

We collected children’s stories with minimal or no punctuation from the web and synthesized them under three scenarios using the End-to-End TTS system described in Section~\ref{sec:end_to_end_tts}. We describe all three different scenarios below

\begin{enumerate}
    \item \textbf{Scenario 1} The story text is passed directly to the TTS system without predicting comma placement. This is referred to as the No Model in the evaluation.
    \item \textbf{Scenario 2} The trained BLSTM model (Section~\ref{subsubsec:phrasing_models_systems_blstm}) is used to punctuate the story text. The punctuated text is then passed to the TTS system for synthesis. This is referred to as the BLSTM Model in the evaluation.
    \item \textbf{Scenario 3} The fine-tuned BERT model (Section~\ref{subsubsec:phrasing_models_systems_blstm_bert}), is used to punctuate the story text. The punctuated text is then fed to the TTS system for synthesis. This is referred to as the BERT Model in the evaluation.
\end{enumerate}

To assess the quality of children's stories synthesized under these three scenarios, we conducted subjective listening tests. These tests were designed as pairwise ABX tasks. Participants were presented with the same story synthesized under two scenarios in a randomized order and asked to identify the preferred version for each story. Additionally, they had the option to select ``no preference" if they found the versions indistinguishable.

Five children's stories were randomly chosen and synthesized under all three scenarios, and used for the evaluation. The evaluation involved 70 post-graduate students from IIIT Hyderabad. All participants were placed in a quiet room and provided with headphones during the evaluation. The evaluation resulted in 350 responses for each pairwise comparison.

\begin{figure*}[!htb]
    \centering
    \includegraphics[scale=0.4]{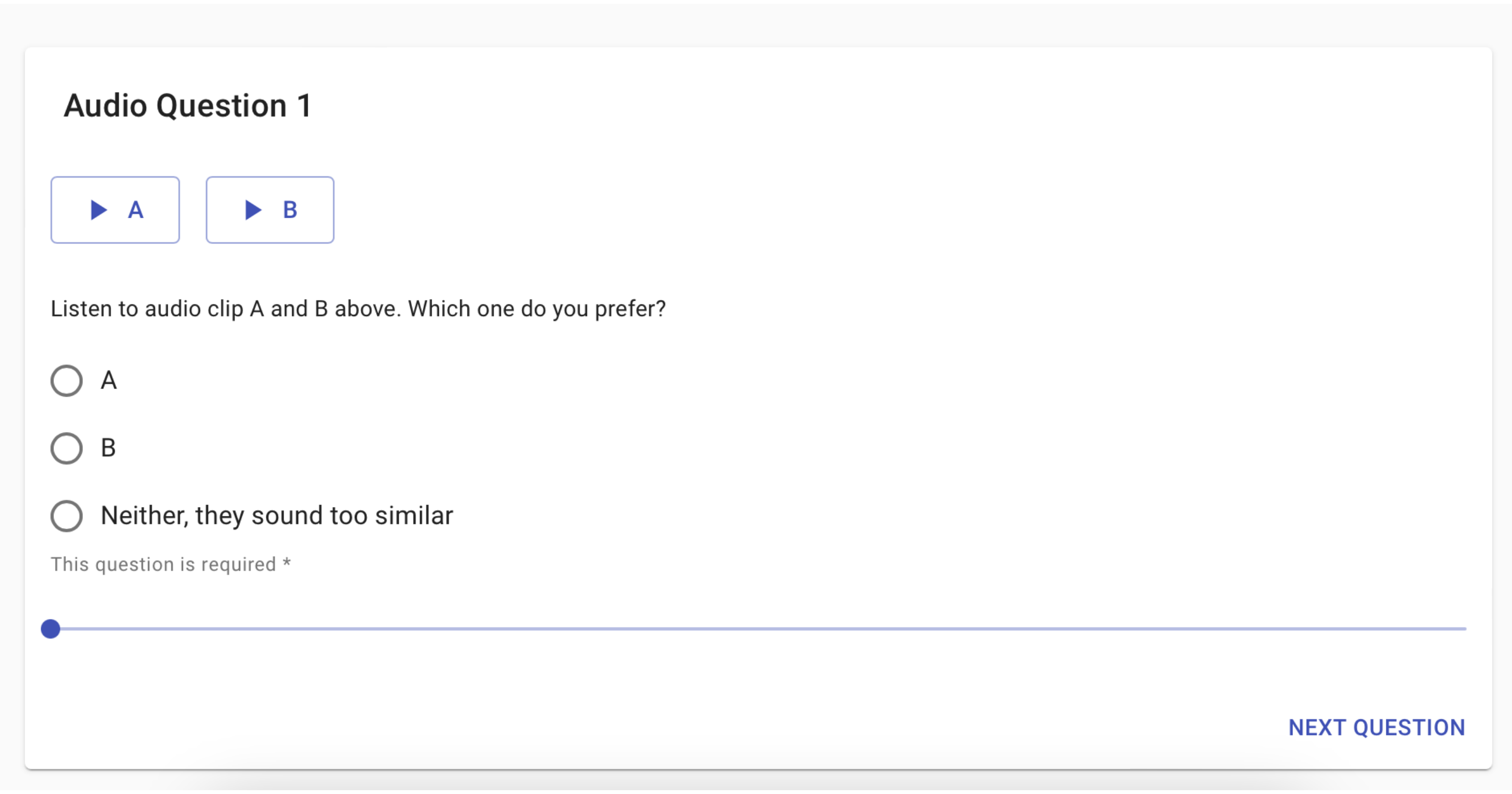}
    \caption{Screen shot of the ABX test interface}
    \label{fig:abx-test-screenshot}
\end{figure*}

The ABX tests were hosted on \url{https://golisten.ucd.ie/}. Figure~\ref{fig:abx-test-screenshot} shows a screenshot of the test interface.

\begin{table*}[!htb]
    \caption{ABX evaluations between No Model, BLSTM Model and BERT Model. *** indicates statistically significant difference at 1\% level as a result of a chi-squared test}
    \centering
    \begin{tabular}{|c|c||c|c|c|c|}\hline
        Model A     & Model B     & Preference A & Preference B & No Preference &     \\ \hline
        No Model    & BLSTM Model & 112          & \textbf{156} & 82            & *** \\ \hline
        No Model    & BERT Model  & 99           & \textbf{183} & 68            & *** \\ \hline
        BLSTM Model & BERT Model  & 111          & \textbf{163} & 76            & *** \\ \hline
    \end{tabular}
    \label{tab:ABXResults}
\end{table*}

Table~\ref{tab:ABXResults} presents the results of pairwise ABX evaluations conducted across the three scenarios. In the first comparison between the No Model and the BLSTM model, 112 responses expressed a preference for the No Model, 156 preferred the BLSTM model, and 82 indicated no preference. Similarly, in the second comparison between the No Model and the BERT model, 99 responses favored the No Model, 183 preferred the BERT model, and 68 had no preference. In the final comparison between the BLSTM and BERT models, 111 responses preferred the BLSTM model, 156 favored the BERT model, and 76 expressed no preference.

These results demonstrate a clear preference for text punctuated using a trained phrasing model (either the BLSTM or BERT model) prior to synthesis, in contrast to text synthesized without prior comma prediction. Furthermore, the comparison between the two trained phrasing models indicates a preference for the BERT model over the BLSTM model.

To assess the statistical significance of the observed differences in preferences, a chi-squared test was performed for each comparison. The results revealed that the differences in preferences were statistically significant at the 1\% significance level, suggesting a meaningful distinction between the models in terms of user preference.

These findings support the utility of incorporating explicit phrasing models within End-to-End TTS systems to enhance listener comprehension. Furthermore, these results in combination with those reported in Section~\ref{subsec:phrasing_models_objective_results} suggest that pre-trained language models, such as BERT, offer potential advantages in phrase break prediction for TTS systems.

\section{Conclusions}
\label{sec:summary_conclusions}

This paper investigated phrase break prediction within End-to-End TTS systems, driven by two key questions: (i) Does incorporating an explicit external phrasing model benefit End-to-End TTS?, and (ii) How can we assess the effectiveness of an external phrasing model?

We evaluated the impact of phrase break prediction models in synthesizing children’s stories. Listener comprehension served as the primary metric to assess the influence of these models on the clarity and understandability of the synthesized speech.

The End-to-End TTS system employed in this work consisted of two components: (i) A Tacotron2 model; with Dynamic Convolutional Attention which modifies the hybrid location sensitive attention mechanism to be purely location based, resulting in better generalization on long utterances, and (ii) A WaveRNN based vocoder; which takes the mel spectrogram predicted by the Tacotron2 model as input and generates a waveform as output. Typically, TTS systems introduce pauses at commas during synthesis. We trained our End-to-End TTS system to adhere to this.

Two phrasing models were investigated in this work: (i) A BLSTM model with task-specific embeddings trained from scratch, and (ii) A BERT model fine-tuned on phrase break prediction. Both models were trained on a multi-speaker English dataset. When presented with unpunctuated text, these models predict comma locations, essentially punctuating the text. The punctuated text is then passed to the TTS system for synthesis.

Unpunctuated children's stories were collected from the web and synthesized under three different scenarios using the End-to-End TTS system, as described in Section~\ref{sec:end_to_end_tts_phrasing}. We conducted subjective listening tests in a pairwise ABX format to evaluate the synthesized stories. The results of the listening tests revealed a clear preference for text punctuated using the trained phrasing models before being synthesized, compared to text directly synthesized without comma prediction. Furthermore, the listening test results indicate a preference for the BERT model over the BLSTM model. These findings support the utility of incorporating explicit phrasing models within End-to-End TTS systems.

\section{Limitations \& future work}

This work primarily focuses on predicting pause locations during speech synthesis using our phrase break models. We acknowledge, as discussed in Section~\ref{sec:phrase_break}, that phrase breaks manifest in speech through various acoustic cues beyond pauses. These cues include relative changes in intonation, syllable duration, and other factors like pre-pausal lengthening, speaking rate, breaths, and glottalization. The task of representing and modeling these non-pausal cues for phrase break prediction remains open for future exploration.

The End-to-End TTS system utilized in this work leverages a Tacotron2 model with Dynamic Convolutional Attention and a WaveRNN vocoder. While this system achieves good performance, recent advancements in speech synthesis have seen the rise of generative models like Vall-E~\cite{DBLP:journals/corr/abs-2302-03540} and SPEAR-TTS~\cite{DBLP:journals/corr/abs-2301-02111}. These models outperform autoregressive models like Tacotron2 in naturalness and expressiveness.  The questions regarding the utility and effectiveness of phrasing models, as investigated in this work, can also be extended to these newer generative models. We posit that exploring the role of phrasing models within these advanced architectures is a promising direction for future research.
\backmatter

\section*{Declarations}

\begin{itemize}
    \item[] \textbf{Funding} No funds, grants or other support was received during this work.
    \item[] \textbf{Conflict of interest} The author has no conflicts or competing interests.
    \item[] \textbf{Ethics approval and consent to participate} Not Applicable
    \item[] \textbf{Consent for publication} The author consents to the publication of this work.
    \item[] \textbf{Data availability}
          \begin{itemize}
              \item The data used to train the phrasing model is publicly available at \url{https://github.com/kan-bayashi/LibriTTSLabel}
              \item The data used to train the End-to-End TTS system is publicly available at \url{https://keithito.com/LJ-Speech-Dataset/}
              \item All samples used in the ABX listening tests described in the paper are available online at \url{https://anandaswarup.github.io/phrase_break_prediction/}
          \end{itemize}
    \item[] \textbf{Materials availability} Not Applicable
    \item[] \textbf{Code availability}
          \begin{itemize}
              \item The code for the phrasing models is available online at \url{https://github.com/anandaswarup/phrase_break_prediction}
              \item The code for the End-to-End TTS system is available online at \url{https://github.com/anandaswarup/RNN-TTS}
          \end{itemize}
\end{itemize}

\bibliography{refs}

\begin{thebibliography}{10}
\providecommand{\url}[1]{{#1}}
\providecommand{\urlprefix}{URL }
\providecommand{\doi}[1]{\url{https://doi.org/#1}}
\bibcommenthead

\bibitem{Skerry-Ryan2018}
R.J. Skerry{-}Ryan, E.~Battenberg, Y.~Xiao, Y.~Wang, D.~Stanton, J.~Shor, R.J.
  Weiss, R.~Clark, R.A. Saurous, \emph{Towards {E}nd-to-{E}nd Prosody Transfer
  for Expressive Speech Synthesis with {T}acotron}, in \emph{Proceedings of the
  35th International Conference on Machine Learning, {ICML} 2018,
  Stockholmsm{\"{a}}ssan, Stockholm, Sweden, July 10-15, 2018},
  \emph{Proceedings of Machine Learning Research}, vol.~80, ed. by J.G. Dy,
  A.~Krause ({PMLR}, 2018), pp. 4700--4709

\bibitem{Wang2018}
Y.~Wang, D.~Stanton, Y.~Zhang, R.J. Skerry{-}Ryan, E.~Battenberg, J.~Shor,
  Y.~Xiao, Y.~Jia, F.~Ren, R.A. Saurous, \emph{Style {T}okens: Unsupervised
  Style Modeling, Control and Transfer in {E}nd-to-{E}nd Speech Synthesis}, in
  \emph{Proceedings of the 35th International Conference on Machine Learning,
  {ICML} 2018, Stockholmsm{\"{a}}ssan, Stockholm, Sweden, July 10-15, 2018},
  \emph{Proceedings of Machine Learning Research}, vol.~80, ed. by J.G. Dy,
  A.~Krause ({PMLR}, 2018), pp. 5167--5176

\bibitem{Zhang2019}
Y.~Zhang, S.~Pan, L.~He, Z.~Ling, \emph{Learning Latent Representations for
  Style Control and Transfer in {E}nd-to-{E}nd Speech Synthesis}, in
  \emph{{IEEE} International Conference on Acoustics, Speech and Signal
  Processing, {ICASSP} 2019, Brighton, United Kingdom, May 12-17, 2019}
  ({IEEE}, 2019), pp. 6945--6949.
\newblock \doi{10.1109/ICASSP.2019.8683623}

\bibitem{Lee2019}
Y.~Lee, T.~Kim, \emph{Robust and Fine-grained Prosody Control of {E}nd-to-{E}nd
  Speech Synthesis}, in \emph{{IEEE} International Conference on Acoustics,
  Speech and Signal Processing, {ICASSP} 2019, Brighton, United Kingdom, May
  12-17, 2019} ({IEEE}, 2019), pp. 5911--5915.
\newblock \doi{10.1109/ICASSP.2019.8683501}

\bibitem{Hsu2019}
W.~Hsu, Y.~Zhang, R.J. Weiss, H.~Zen, Y.~Wu, Y.~Wang, Y.~Cao, Y.~Jia, Z.~Chen,
  J.~Shen, P.~Nguyen, R.~Pang, \emph{Hierarchical Generative Modeling for
  Controllable Speech Synthesis}, in \emph{7th International Conference on
  Learning Representations, {ICLR} 2019, New Orleans, LA, USA, May 6-9, 2019}
  (OpenReview.net, 2019)

\bibitem{Battenberg2019}
E.~Battenberg, S.~Mariooryad, D.~Stanton, R.J. Skerry{-}Ryan, M.~Shannon,
  D.~Kao, T.~Bagby, Effective use of variational embedding capacity in
  expressive end-to-end speech synthesis.
\newblock CoRR \textbf{abs/1906.03402} (2019)

\bibitem{Aggarwal2020}
V.~Aggarwal, M.~Cotescu, N.~Prateek, J.~Lorenzo{-}Trueba, R.~Barra{-}Chicote,
  \emph{Using {VAE}s and {N}ormalizing {F}lows for {O}ne-{S}hot
  {T}ext-To-{S}peech {S}ynthesis of Expressive Speech}, in \emph{2020 {IEEE}
  International Conference on Acoustics, Speech and Signal Processing, {ICASSP}
  2020, Barcelona, Spain, May 4-8, 2020} ({IEEE}, 2020), pp. 6179--6183.
\newblock \doi{10.1109/ICASSP40776.2020.9053678}

\bibitem{Valle2020}
R.~Valle, J.~Li, R.~Prenger, B.~Catanzaro, \emph{Mellotron: {M}ultispeaker
  {E}xpressive {V}oice {S}ynthesis by Conditioning on {R}hythm, {P}itch and
  {G}lobal {S}tyle {T}okens}, in \emph{2020 {IEEE} International Conference on
  Acoustics, Speech and Signal Processing, {ICASSP} 2020, Barcelona, Spain, May
  4-8, 2020} ({IEEE}, 2020), pp. 6189--6193.
\newblock \doi{10.1109/ICASSP40776.2020.9054556}

\bibitem{Hu2020}
T.~Hu, A.~Shrivastava, O.~Tuzel, C.~Dhir, \emph{Unsupervised {S}tyle and
  {C}ontent {S}eparation by {M}inimizing {M}utual {I}nformation for {S}peech
  {S}ynthesis}, in \emph{2020 {IEEE} International Conference on Acoustics,
  Speech and Signal Processing, {ICASSP} 2020, Barcelona, Spain, May 4-8, 2020}
  ({IEEE}, 2020), pp. 3267--3271.
\newblock \doi{10.1109/ICASSP40776.2020.9054591}

\bibitem{DBLP:journals/spl/LiuSBGL20}
R.~Liu, B.~Sisman, F.~Bao, G.~Gao, H.~Li, Modeling prosodic phrasing with
  multi-task learning in tacotron-based {TTS}.
\newblock {IEEE} Signal Process. Lett. \textbf{27}, 1470--1474 (2020).
\newblock \doi{10.1109/LSP.2020.3016564}.
\newblock \urlprefix\url{https://doi.org/10.1109/LSP.2020.3016564}

\bibitem{DBLP:conf/icassp/XiaoHMS20}
Y.~Xiao, L.~He, H.~Ming, F.K. Soong, \emph{Improving Prosody with Linguistic
  and Bert Derived Features in Multi-Speaker Based Mandarin Chinese Neural
  {TTS}}, in \emph{2020 {IEEE} International Conference on Acoustics, Speech
  and Signal Processing, {ICASSP} 2020, Barcelona, Spain, May 4-8, 2020}
  ({IEEE}, 2020), pp. 6704--6708.
\newblock \doi{10.1109/ICASSP40776.2020.9054337}.
\newblock \urlprefix\url{https://doi.org/10.1109/ICASSP40776.2020.9054337}

\bibitem{Silverman1992}
K.E.A. Silverman, M.E. Beckman, J.F. Pitrelli, M.~Ostendorf, C.W. Wightman,
  P.~Price, J.B. Pierrehumbert, J.~Hirschberg, \emph{{ToBI}: a standard for
  labeling {E}nglish prosody.}, in \emph{ICSLP} (ISCA, 1992)

\bibitem{taylor1998}
P.~Taylor, A.W. Black, Assigning phrase breaks from part-of-speech sequences.
\newblock Comput. Speech Lang. \textbf{12}(2), 99--117 (1998).
\newblock \doi{10.1006/csla.1998.0041}

\bibitem{kishoreSSW7}
K.~Prahallad, E.V. Raghavendra, A.W. Black, \emph{Learning speaker-specific
  phrase breaks for text-to-speech systems}, in \emph{Proceedings of 7th {ISCA}
  {S}peech {S}ynthesis {W}orkshop ({SSW7})} (Kyoto, Japan, 2010), pp. 148--153

\bibitem{kishore2010}
K.~Prahallad, E.V. Raghavendra, A.W. Black, \emph{Semi-supervised learning of
  acoustic driven prosodic phrase breaks for text-to-speech systems}, in
  \emph{Proceedings of 5th International Conference on Speech Prosody (Speech
  Prosody 2010)} (Chicago, Illinois, 2010)

\bibitem{Hirschberg1993}
J.~Hirschberg, Pitch accent in context: Predicting intonational prominence from
  text.
\newblock Artif. Intell. \textbf{63}(1-2), 305--340 (1993).
\newblock \doi{10.1016/0004-3702(93)90020-C}

\bibitem{Ross1996}
K.N. Ross, M.~Ostendorf, Prediction of abstract prosodic labels for speech
  synthesis.
\newblock Comput. Speech Lang. \textbf{10}(3), 155--185 (1996).
\newblock \doi{10.1006/csla.1996.0010}

\bibitem{Santen1994}
J.P.H. van Santen, Assignment of segmental duration in text-to-speech
  synthesis.
\newblock Comput. Speech Lang. \textbf{8}(2), 95--128 (1994).
\newblock \doi{10.1006/csla.1994.1005}

\bibitem{wightman1992}
C.~Wightman, S.~Shattuck-Hufnagel, M.~Ostendorf, P.J. Price, Segmental
  durations in the vicinity of prosodic phrase boundaries.
\newblock Journal Acoustical Society of America \textbf{91}(3), 1707--1717
  (1992)

\bibitem{redi2001}
L.~Redi, S.~Shattuck-Hufnagel, Variation in realization of glottalization in
  normal speakers.
\newblock Journal of Phonetics \textbf{29}, 407--429 (2001)

\bibitem{kim2006}
H.~Kim, T.~Yoon, J.~Cole, M.~Hasegawa-Johnson, \emph{Acoustic differentiation
  of {L}- and {L-L}\% in switchboard and radio news speech}, in
  \emph{Proceedings of Speech Prosody} (Dresden, 2006)

\bibitem{parlikar2013}
A.~Parlikar, A.W. Black, \emph{Minimum error rate training for phrasing in
  Speech Synthesis}, in \emph{Procedings of 8th ISCA Speech Synthesis Workshop
  (SSW8)} (Barcelona, Spain, 2013), pp. 13--16

\bibitem{wattsBlizzard2013}
O.~Watts, A.~Stan, Y.~Mamiya, A.~Suni, J.M. Burgos, J.M. Montero, \emph{The
  {Simple4All} entry to the Blizzard Challenge 2013}, in \emph{Proceedings of
  the 2013 Blizzard Challenge Workshop} (2013)

\bibitem{anandaswarup2013}
A.~Vadapalli, P.~Bhaskararao, K.~Prahallad, \emph{Significance of word-terminal
  syllables for prediction of phrase breaks in {T}ext-to-{S}peech systems in
  {I}ndian Languages}, in \emph{Proceedings of 8th ISCA Speech Synthesis
  Workshop (SSW8)} (Barcelona, Spain, 2013), pp. 189--194

\bibitem{watts2011}
O.~Watts, J.~Yamagishi, S.~King, \emph{Unsupervised continuous-valued word
  features for phrase-break prediction without a part-of-speech tagger}, in
  \emph{Proceedings of Interspeech} (Florence, Italy, 2011), pp. 2157--2160

\bibitem{anandaswarupIS14}
A.~Vadapalli, K.~Prahallad, \emph{Learning continuous-valued word
  representations for phrase break prediction}, in \emph{{INTERSPEECH} 2014,
  15th Annual Conference of the International Speech Communication Association,
  Singapore, September 14-18, 2014} ({ISCA}, 2014), pp. 41--45

\bibitem{wattsICASSP2014}
O.~Watts, S.~Gangireddy, J.~Yamagishi, S.~King, S.~Renals, A.~Stan, M.~Giurgiu,
  \emph{Neural Net Word Representations for Phrase-Break Prediction Without a
  Part of Speech Tagger}, in \emph{Proceedings IEEE International Conference on
  Acoustics, Speech, and Signal Processing (ICASSP)} (Florence, Italy, 2014),
  pp. 2599--2603

\bibitem{anandaswarupIS16}
A.~Vadapalli, S.V. Gangashetty, \emph{An Investigation of Recurrent Neural
  Network Architectures Using Word Embeddings for Phrase Break Prediction}, in
  \emph{{INTERSPEECH} 2016, 17th Annual Conference of the International Speech
  Communication Association, San Francisco, CA, USA, September 8-12, 2016}
  ({ISCA}, 2016), pp. 2308--2312.
\newblock \doi{10.21437/Interspeech.2016-885}

\bibitem{parlikar2011}
A.~Parlikar, A.W. Black, \emph{A Grammar Based Approach to Style Specific
  Phrase Prediction}, in \emph{Proceedings of Interspeech} (Florence, Italy,
  2011), pp. 2149--2152

\bibitem{wang1992}
M.~Wang, J.~Hirschberg, Automatic classification of intonational phrase
  boundaries.
\newblock Computer Speech and Language \textbf{6}, 175--196 (1992)

\bibitem{navas2008}
E.~Navas, I.~Hernez, I.~Sainz, Evaluation of automatic break insertion for an
  agglutinative and inflected language.
\newblock Speech Communication \textbf{50}(11-12), 888--899 (2008)

\bibitem{schmid2004}
H.~Schmid, M.~Atterer, \emph{New statistical methods for phrase break
  prediction}, in \emph{Proceedings of 20th {I}nternational conference on
  {C}omputational {L}inguistics,{COLING '04}} (Geneva, Switzerland, 2004)

\bibitem{bonafonte2004}
A.~Bonafonte, P.~Ag{\"u}ero, \emph{Phrase break prediction using a finite state
  transducer}, in \emph{Proceedings of 11th International Workshop on Advances
  in Speech Technology} (2004)

\bibitem{busser2001}
B.~Busser, W.~Daelemans, A.~van~den Bosch, \emph{Predicting phrase breaks with
  memory-based learning}, in \emph{Proceedings of 4th {ISCA} Speech Synthesis
  Workshop} (2001)

\bibitem{parlikar2012}
A.~Parlikar, A.W. Black, \emph{Data-Driven Phrasing for Speech Synthesis in
  Low-Resource Langauges}, in \emph{Proceedings of IEEE International
  Conference on Acoustics, Speech and Signal Processing} (Kyoto, Japan, 2012)

\bibitem{hema2004}
N.S. Krishna, H.A. Murthy, \emph{A new prosodic phrasing model for {I}ndian
  language {T}elugu}, in \emph{{INTERSPEECH-2004-ICSLP}}, vol.~1 (2004), pp.
  793--796

\bibitem{wattsThesis}
O.~Watts, Unsupervised learning for text-to-speech synthesis.
\newblock Ph.D. thesis, {U}niversity of {E}dinburgh (2012)

\bibitem{DBLP:conf/interspeech/KlimkovNMPBMD17}
V.~Klimkov, A.~Nadolski, A.~Moinet, B.~Putrycz, R.~Barra{-}Chicote, T.~Merritt,
  T.~Drugman, \emph{Phrase Break Prediction for Long-Form Reading {TTS:}
  Exploiting Text Structure Information}, in \emph{Interspeech 2017, 18th
  Annual Conference of the International Speech Communication Association,
  Stockholm, Sweden, August 20-24, 2017}, ed. by F.~Lacerda ({ISCA}, 2017), pp.
  1064--1068.
\newblock \doi{10.21437/INTERSPEECH.2017-419}.
\newblock \urlprefix\url{https://doi.org/10.21437/Interspeech.2017-419}

\bibitem{DBLP:conf/naacl/DevlinCLT19}
J.~Devlin, M.~Chang, K.~Lee, K.~Toutanova, \emph{{BERT:} Pre-training of Deep
  Bidirectional Transformers for Language Understanding}, in \emph{Proceedings
  of the 2019 Conference of the North American Chapter of the Association for
  Computational Linguistics: Human Language Technologies, {NAACL-HLT} 2019,
  Minneapolis, MN, USA, June 2-7, 2019, Volume 1 (Long and Short Papers)}, ed.
  by J.~Burstein, C.~Doran, T.~Solorio (Association for Computational
  Linguistics, 2019), pp. 4171--4186.
\newblock \doi{10.18653/V1/N19-1423}.
\newblock \urlprefix\url{https://doi.org/10.18653/v1/n19-1423}

\bibitem{DBLP:conf/interspeech/FutamataPYT21}
K.~Futamata, B.~Park, R.~Yamamoto, K.~Tachibana, \emph{Phrase Break Prediction
  with Bidirectional Encoder Representations in Japanese Text-to-Speech
  Synthesis}, in \emph{22nd Annual Conference of the International Speech
  Communication Association, Interspeech 2021, Brno, Czechia, August 30 -
  September 3, 2021}, ed. by H.~Hermansky, H.~Cernock{\'{y}}, L.~Burget,
  L.~Lamel, O.~Scharenborg, P.~Motl{\'{\i}}cek ({ISCA}, 2021), pp. 3126--3130.
\newblock \doi{10.21437/INTERSPEECH.2021-252}.
\newblock \urlprefix\url{https://doi.org/10.21437/Interspeech.2021-252}

\bibitem{DBLP:conf/interspeech/ZenDCZWJCW19}
H.~Zen, V.~Dang, R.~Clark, Y.~Zhang, R.J. Weiss, Y.~Jia, Z.~Chen, Y.~Wu,
  \emph{Libri{TTS}: {A} Corpus Derived from LibriSpeech for Text-to-Speech}, in
  \emph{Interspeech 2019, 20th Annual Conference of the International Speech
  Communication Association, Graz, Austria, 15-19 September 2019}, ed. by
  G.~Kubin, Z.~Kacic ({ISCA}, 2019), pp. 1526--1530.
\newblock \doi{10.21437/Interspeech.2019-2441}.
\newblock \urlprefix\url{https://doi.org/10.21437/Interspeech.2019-2441}

\bibitem{DBLP:conf/interspeech/McAuliffeSM0S17}
M.~McAuliffe, M.~Socolof, S.~Mihuc, M.~Wagner, M.~Sonderegger, \emph{Montreal
  Forced Aligner: Trainable Text-Speech Alignment Using Kaldi}, in
  \emph{Interspeech 2017, 18th Annual Conference of the International Speech
  Communication Association, Stockholm, Sweden, August 20-24, 2017}, ed. by
  F.~Lacerda ({ISCA}, 2017), pp. 498--502.
\newblock \doi{10.21437/INTERSPEECH.2017-1386}.
\newblock \urlprefix\url{https://doi.org/10.21437/Interspeech.2017-1386}

\bibitem{DBLP:journals/corr/KingmaB14}
D.P. Kingma, J.~Ba, \emph{Adam: {A} Method for Stochastic Optimization}, in
  \emph{3rd International Conference on Learning Representations, {ICLR} 2015,
  San Diego, CA, USA, May 7-9, 2015, Conference Track Proceedings}, ed. by
  Y.~Bengio, Y.~LeCun (2015).
\newblock \urlprefix\url{http://arxiv.org/abs/1412.6980}

\bibitem{FMeasure}
C.J. van Rijsbergen, \emph{Information Retrieval} (Butterworth, 1979)

\bibitem{DBLP:conf/icassp/BattenbergSMSKS20}
E.~Battenberg, R.J. Skerry{-}Ryan, S.~Mariooryad, D.~Stanton, D.~Kao,
  M.~Shannon, T.~Bagby, \emph{Location-Relative Attention Mechanisms for Robust
  Long-Form Speech Synthesis}, in \emph{2020 {IEEE} International Conference on
  Acoustics, Speech and Signal Processing, {ICASSP} 2020, Barcelona, Spain, May
  4-8, 2020} ({IEEE}, 2020), pp. 6194--6198.
\newblock \doi{10.1109/ICASSP40776.2020.9054106}.
\newblock \urlprefix\url{https://doi.org/10.1109/ICASSP40776.2020.9054106}

\bibitem{DBLP:conf/icml/KalchbrennerESN18}
N.~Kalchbrenner, E.~Elsen, K.~Simonyan, S.~Noury, N.~Casagrande, E.~Lockhart,
  F.~Stimberg, A.~van~den Oord, S.~Dieleman, K.~Kavukcuoglu, \emph{Efficient
  Neural Audio Synthesis}, in \emph{Proceedings of the 35th International
  Conference on Machine Learning, {ICML} 2018, Stockholmsm{\"{a}}ssan,
  Stockholm, Sweden, July 10-15, 2018}, \emph{Proceedings of Machine Learning
  Research}, vol.~80, ed. by J.G. Dy, A.~Krause ({PMLR}, 2018), pp. 2415--2424.
\newblock \urlprefix\url{http://proceedings.mlr.press/v80/kalchbrenner18a.html}

\bibitem{DBLP:conf/interspeech/Lorenzo-TruebaD19}
J.~Lorenzo{-}Trueba, T.~Drugman, J.~Latorre, T.~Merritt, B.~Putrycz,
  R.~Barra{-}Chicote, A.~Moinet, V.~Aggarwal, \emph{Towards Achieving Robust
  Universal Neural Vocoding}, in \emph{Interspeech 2019, 20th Annual Conference
  of the International Speech Communication Association, Graz, Austria, 15-19
  September 2019}, ed. by G.~Kubin, Z.~Kacic ({ISCA}, 2019), pp. 181--185.
\newblock \doi{10.21437/Interspeech.2019-1424}.
\newblock \urlprefix\url{https://doi.org/10.21437/Interspeech.2019-1424}

\bibitem{ljspeech17}
K.~Ito, L.~Johnson.
\newblock The {LJ} speech dataset.
\newblock \url{https://keithito.com/LJ-Speech-Dataset/} (2017)

\bibitem{DBLP:journals/corr/abs-2302-03540}
E.~Kharitonov, D.~Vincent, Z.~Borsos, R.~Marinier, S.~Girgin, O.~Pietquin,
  M.~Sharifi, M.~Tagliasacchi, N.~Zeghidour, Speak, read and prompt:
  High-fidelity text-to-speech with minimal supervision.
\newblock CoRR \textbf{abs/2302.03540} (2023).
\newblock \doi{10.48550/arXiv.2302.03540}.
\newblock \urlprefix\url{https://doi.org/10.48550/arXiv.2302.03540}.
\newblock {\href{https://arxiv.org/abs/2302.03540}{{2302.03540}}}

\bibitem{DBLP:journals/corr/abs-2301-02111}
C.~Wang, S.~Chen, Y.~Wu, Z.~Zhang, L.~Zhou, S.~Liu, Z.~Chen, Y.~Liu, H.~Wang,
  J.~Li, L.~He, S.~Zhao, F.~Wei, Neural codec language models are zero-shot
  text to speech synthesizers.
\newblock CoRR \textbf{abs/2301.02111} (2023).
\newblock \doi{10.48550/arXiv.2301.02111}.
\newblock \urlprefix\url{https://doi.org/10.48550/arXiv.2301.02111}.
\newblock {\href{https://arxiv.org/abs/2301.02111}{{2301.02111}}}

\end{thebibliography}

\end{document}